\documentclass[aps,rpl,preprint,groupedaddress]{revtex4}
\begin{document}

\title{Teleportation and Dense Coding \\ with Genuine Multipartite Entanglement}
\author{Ye Yeo}
\affiliation{Department of Physics, National University of Singapore, 10 Kent Ridge Crescent, Singapore 119260, Singapore}

\author{Wee Kang Chua}
\affiliation{Department of Physics, National University of Singapore, 10 Kent Ridge Crescent, Singapore 119260, Singapore}

\begin{abstract}
We present an explicit protocol ${\cal E}_0$ for faithfully teleporting an arbitrary two-qubit state via a genunie four-qubit entangled state.  By construction, our four-partite state is not reducible to a pair of Bell states.  Its properties are compared and contrasted with those of the four-party GHZ and W states.  We also give a dense coding scheme ${\cal D}_0$ involving our state as a shared resource of entanglement.  Both ${\cal D}_0$ and ${\cal E}_0$ indicate that our four-qubit state is a likely candidate for the genunine four-partite analogue to a Bell state.
\end{abstract}

\maketitle

Quantum teleportation, the disembodied transport of quantum states between subsystems through a classical communication channel requiring a shared resource of entanglement, is one of the most profound results of quantum information theory \cite{Nielsen}.  Bennett {\em et al.} \cite{Bennett1} are the first to show how quantum entanglement can assist in the teleportation of an intact quantum state
\begin{equation}
|\psi\rangle_{A_1} = a|0\rangle_{A_1} + b|1\rangle_{A_1},
\end{equation}
with $a, b \in {\cal C}^1$ and $|a|^2 + |b|^2 = 1$, from one place to another, by a sender, Alice, who knows neither the state $|\psi\rangle_{A_1}$ to be teleported nor the location of the intended receiver, Bob.  In their standard teleportation protocol ${\cal T}_0$, Alice and Bob share {\em a priori} a pair of particles, $A_2$ and $B$, in a maximally entangled Bell state, say
\begin{equation}
|\Psi^0_{Bell}\rangle_{A_2B} \equiv \frac{1}{\sqrt{2}}(|00\rangle_{A_2B} + |11\rangle_{A_2B}).
\end{equation}
Teleportation firmly establishes the practical basis for considering the maximally entangled Bell states as basic units, upon which bipartite entanglement can be quantitatively expressed in terms of.  Indeed, quantities like the concurrence \cite{Hill, Wootters} and fully entangled fraction \cite{Horodecki} have their roots in these states.

The teleportation of an arbitrary two-qubit state:
\begin{equation}
|\Psi\rangle_{A_1A_2} = a|00\rangle_{A_1A_2} + b|01\rangle_{A_1A_2} + c|10\rangle_{A_1A_2} + d|11\rangle_{A_1A_2},
\end{equation}
with $a, b, c, d \in {\cal C}^1$ and $|a|^2 + |b|^2 + |c|^2 + |d|^2 = 1$; had been studied by Lee {\em et al.} \cite{Lee} and recently by Rigolin \cite{Rigolin}.  Whereas Lee {\em et al.} did not explicitly construct a protocol, the 16 $G$ states defined by Rigolin in his protocol: $|G^{ij}\rangle_{A_3A_4B_1B_2} \equiv [(\sigma^i_{A_3} \otimes \sigma^j_{A_4}) \otimes I_{B_1B_2}]|G^{00}\rangle_{A_3A_4B_1B_2}$, with
\begin{eqnarray}
|G^{00}\rangle_{A_3A_4B_1B_2} & \equiv & \frac{1}{2}
(|0000\rangle_{A_3A_4B_1B_2} + |0101\rangle_{A_3A_4B_1B_2} + |1010\rangle_{A_3A_4B_1B_2} + |1111\rangle_{A_3A_4B_1B_2}), \nonumber \\
& = & |\Psi^0_{Bell}\rangle_{A_3B_1} \otimes |\Psi^0_{Bell}\rangle_{A_4B_2},
\end{eqnarray}
were actually tensor products of two Bell states.  This fact had been highlighted in Ref.\cite{Lee}.  In this paper, we give an explicit protocol ${\cal E}_0$ for faithfully teleporting arbitrary two-qubit states employing genuine four-qubit entangled states $|\bar{\chi}^{00}\rangle$ [Eq.(9)], and especially $|\chi^{00}\rangle$ [Eq.(18)].  This is an important consideration because $|\chi^{00}\rangle$, in addition to $|\Psi^0_{Bell}\rangle \otimes |\Psi^0_{Bell}\rangle$, could be a likely candidate for the genuine four-partite analogue to $|\Psi^0_{Bell}\rangle$.  Since our work is motivated in part by ${\cal T}_0$, we briefly describe it below before presenting our protocol ${\cal E}_0$.  This is followed by a detailed analysis on the entanglement properties of $|\bar{\chi}^{00}\rangle$ and $|\chi^{00}\rangle$, where we compare and contrast with those of the four-party GHZ and W states.  Before concluding, we give a dense coding scheme ${\cal D}_0$ using $|\chi^{00}\rangle$ as the shared resource of entanglement.

In ${\cal T}_0$, the initial complete state of the three particles, $A_1$, $A_2$ and $B$, is a pure product state,
\begin{equation}
|\psi\rangle_{A_1}\langle\psi| \otimes |\Psi^0_{Bell}\rangle_{A_2B}\langle\Psi^0_{Bell}|,
\end{equation}
involving neither classical correlation nor quantum entanglement between particle $A_1$ and the maximally entangled pair $A_2B$.  Alice cleanly divides the full information encoded in $|\psi\rangle_{A_1}$ into two parts, transmitting first the purely nonclassical part via the quantum channel $|\Psi^0_{Bell}\rangle_{A_2B}$, by performing a complete von Neumann measurement in the Bell basis:
\begin{equation}
|\Psi^i_{Bell}\rangle_{A_1A_2} = (\sigma^i_{A_1} \otimes \sigma^0_{A_2})|\Psi^0_{Bell}\rangle_{A_1A_2}\ (i = 0, 1, 2, 3),
\end{equation}
on the joint system consisting of particles $A_1$ and $A_2$.  Here, $\sigma^0 = I_2$ is the two-dimensional identity and $\sigma^i (i = 1, 2, 3)$ are the Pauli matrices.  We emphasize that it is a consequence of the fact that $|\Psi^0_{Bell}\rangle_{A_1A_2}$ is maximally entangled, that the $|\Psi^i_{Bell}\rangle_{A_1A_2}$'s are obtainable from $|\Psi^0_{Bell}\rangle_{A_1A_2}$ by appropriate local one-particle Pauli rotation.  The density operator of Bob's qubit $\rho^i_B$ conditioned on Alice's Bell measurement outcome $i$ is
\begin{eqnarray}
& & \frac{1}{p_i}{\rm tr}_{A_1A_2}[
(|\psi\rangle_{A_1}\langle\psi| \otimes |\Psi^0_{Bell}\rangle_{A_2B}\langle\Psi^0_{Bell}|)
(|\Psi^i_{Bell}\rangle_{A_1A_2}\langle\Psi^i_{Bell}| \otimes I_B)] \nonumber \\
& = & \frac{1}{p_i}
{_{A_1A_2}}\langle\Psi^i_{Bell}|(|\psi\rangle_{A_1} \otimes |\Psi^0_{Bell}\rangle_{A_2B}) \times
({_{A_1}}\langle\psi| \otimes {_{A_2B}}\langle\Psi^0_{Bell}|)|\Psi^i_{Bell}\rangle_{A_1A_2} \nonumber \\
& = & \frac{1}{p_i}
{_{A_1A_2}}\langle\Psi^0_{Bell}|(\sigma^i_{A_1}|\psi\rangle_{A_1} \otimes |\Psi^0_{Bell}\rangle_{A_2B}) \times
({_{A_1}}\langle\psi|\sigma^i_{A_1} \otimes {_{A_2B}}\langle\Psi^0_{Bell}|)|\Psi^0_{Bell}\rangle_{A_1A_2} \nonumber \\
& = & \frac{1}{4p_i}\sigma^i_B|\psi\rangle_B\langle\psi|\sigma^i_B,
\end{eqnarray}
where
$$
p_i = {\rm tr}[(|\psi\rangle_{A_1}\langle\psi| \otimes |\Psi^0_{Bell}\rangle_{A_2B}\langle\Psi^0_{Bell}|)
(|\Psi^i_{Bell}\rangle_{A_1A_2}\langle\Psi^i_{Bell}| \otimes I_B)] = \frac{1}{4}.
$$
It follows that, regardless of the unknown state $|\psi\rangle_{A_1}$, the four measurement outcomes are equally likely.  Alice gains no information about the state $|\psi\rangle_{A_1}$ from her measurement.  She is left with particles $A_1$ and $A_2$ in some maximally entangled Bell state, without any trace of the original $|\psi\rangle_{A_1}$.  The outcome of Alice's measurement constitutes the second purely classical part of the full information encoded in $|\psi\rangle_{A_1}$.  She communicates this two bits of information via a classical channel, after which Bob applies the required Pauli rotation to transform the state of his particle $B$ into an accurate replica of the original state of Alice's particle $A_1$.  Eq.(7) follows from, and the success of ${\cal T}_0$ is guaranteed by, the following identity.  For the maximally entangled state Eq.(2), we have \cite{Braunstein}
\begin{eqnarray}
{_{A_1A_2}}\langle\Psi^0_{Bell}|\Psi^0_{Bell}\rangle_{A_2B} 
& = & \frac{1}{2}\sum^1_{i, j = 0}({_{A_1}}\langle i| \otimes {_{A_2}}\langle i|)(|j\rangle_{A_2} \otimes |j\rangle_B) \nonumber \\
& = & \frac{1}{2}\sum^1_{i = 0} |i\rangle_B \times {_{A_1}}\langle i|.
\end{eqnarray}

Our protocol ${\cal E}_0$ is motivated in particular by Eqs.(4) and (8).  To avoid our four-qubit entangled channel from being reducible to a tensor product of two Bell states, and to ensure the success of faithfully teleporting any arbitrary two-qubit state, Alice and Bob share {\em a priori} two pairs of particles, $A_3A_4$ and $B_1B_2$, in the state
\begin{equation}
|\bar{\chi}^{00}\rangle_{A_3A_4B_1B_2} \equiv \frac{1}{2}\sum^3_{J = 0}|J\rangle_{A_3A_4} \otimes |J'\rangle_{B_1B_2}.
\end{equation}
The $|J\rangle$'s constitute an orthonormal basis, and explicitly 
\begin{eqnarray}
|0\rangle & = & \cos\theta_1 |00\rangle + \sin\theta_1|11\rangle, \nonumber \\
|1\rangle & = & \cos\phi_1   |01\rangle + \sin\phi_1  |10\rangle, \nonumber \\
|2\rangle & = & -\sin\phi_1  |01\rangle + \cos\phi_1  |10\rangle, \nonumber \\
|3\rangle & = & -\sin\theta_1|00\rangle + \cos\theta_1|11\rangle.
\end{eqnarray}
The $|J'\rangle$'s constitute another orthonormal basis:
\begin{eqnarray}
|0'\rangle & = & \cos\theta_2 |00\rangle + \sin\theta_2|11\rangle, \nonumber \\
|1'\rangle & = & \sin\phi_2   |01\rangle + \cos\phi_2  |10\rangle, \nonumber \\
|2'\rangle & = & \cos\phi_2   |01\rangle - \sin\phi_2  |10\rangle, \nonumber \\
|3'\rangle & = & -\sin\theta_2|00\rangle + \cos\theta_2|11\rangle.
\end{eqnarray}
Here, $0 < \theta_1,\ \theta_2,\ \phi_1,\ \phi_2 < \pi/2$, and we demand that $\theta_1 \not= \theta_2$, $\phi_1 \not= \phi_2$.  In particular, we may express
\begin{equation}
|\Psi\rangle_{A_1A_2} = \sum^3_{J = 0}\alpha_J|J'\rangle_{A_1A_2},
\end{equation}
with $\alpha_J \in {\cal C}^1$ and $\sum^3_{J = 1}|\alpha_J|^2 = 1$.  By virtue of the fact that, between $A_3A_4$ and $B_1B_2$, $|\bar{\chi}^{00}\rangle_{A_3A_4B_1B_2}$ is a maximally entangled state [compare with Eq.(2)], we may construct the following basis of 16 orthonormal states [similar to Eq.(16)]:
\begin{eqnarray}
|\bar{\Pi}^{00}\rangle_{A_1A_2A_3A_4} 
& \equiv & \frac{1}{2}\sum^3_{K = 0}|K'\rangle_{A_1A_2} \otimes |K\rangle_{A_3A_4} \nonumber \\
|\bar{\Pi}^{ij}\rangle_{A_1A_2A_3A_4}
& = & [(\sigma^i_{A_1} \otimes \sigma^j_{A_2}) \otimes I_{A_3A_4}]|\bar{\Pi}^{00}\rangle_{A_1A_2A_3A_4}.
\end{eqnarray}
If Alice performs a complete projective measurement jointly on $A_1A_2A_3A_4$ in the above basis with the measurement outcome $ij$, then Bob's pair of particles $B_1B_2$ will be in the state
\begin{eqnarray}
& & \frac{1}{\sqrt{p_{ij}}}{_{A_1A_2A_3A_4}}\langle\bar{\Pi}^{ij}|
(|\Psi\rangle_{A_1A_2} \otimes |\bar{\chi}^{00}\rangle_{A_3A_4B_1B_2}) \nonumber \\
& = & \frac{1}{\sqrt{p_{ij}}}{_{A_1A_2A_3A_4}}\langle\bar{\Pi}^{00}|
[(\sigma^i_{A_1} \otimes \sigma^j_{A_2})|\Psi\rangle_{A_1A_2} \otimes |\bar{\chi}^{00}\rangle_{A_3A_4B_1B_2}] \nonumber \\
& = & \frac{1}{4\sqrt{p_{ij}}}(\sigma^i_{B_1} \otimes \sigma^j_{B_2})|\Psi\rangle_{B_1B_2}.
\end{eqnarray}
Here, $|\Psi\rangle_{A_1A_2} \otimes |\bar{\chi}^{00}\rangle_{A_3A_4B_1B_2}$ is the initial complete state of the six particles, $A_1$, $A_2$, $A_3$, $A_4$, $B_1$ and $B_2$.  Eq.(14) is the analogue of Eq.(7).  And, as in ${\cal T}_0$, the success of ${\cal E}_0$ is guaranteed by the following identity:
\begin{eqnarray}
{_{A_1A_2A_3A_4}}\langle\bar{\Pi}^{00}|\bar{\chi}^{00}\rangle_{A_3A_4B_1B_2}
& = & \frac{1}{4}\sum^3_{J, K = 0}
({_{A_1A_2}}\langle K'| \otimes {_{A_3A_4}}\langle K|)(|J\rangle_{A_3A_4} \otimes |J'\rangle_{B_1B_2}) \nonumber \\
& = & \frac{1}{4}\sum^3_{J = 0}|J'\rangle_{B_1B_2} \times {_{A_1A_2}}\langle J'|.
\end{eqnarray}
Clearly, $p_{ij} = 1/16$ and Bob will always succeed in recovering an exact replica of the original state Eq.(12) of Alice's particles $A_1A_2$, upon receiving four bits of classical information about her measurement result.

Now, let us consider the entanglement properties of
\begin{eqnarray}
|\bar{\chi}^{00}\rangle_{A_3A_4B_1B_2} = \frac{1}{2} & [ &
      \cos(\theta_1 - \theta_2)(|0000\rangle_{A_3A_4B_1B_2} + |1111\rangle_{A_3A_4B_1B_2}) \nonumber \\
& - & \sin(\theta_1 - \theta_2)(|0011\rangle_{A_3A_4B_1B_2} - |1100\rangle_{A_3A_4B_1B_2}) \nonumber \\
& - & \sin(\phi_1 - \phi_2)    (|0101\rangle_{A_3A_4B_1B_2} - |1010\rangle_{A_3A_4B_1B_2}) \nonumber \\
& + & \cos(\phi_1 - \phi_2)    (|0110\rangle_{A_3A_4B_1B_2} + |1001\rangle_{A_3A_4B_1B_2})].
\end{eqnarray}
By inspection, we would also have maximum entanglement between $A_3B_1$ and $A_4B_2$ if we demand that $\phi_1 - \phi_2 = \theta_1 - \theta_2$.  In this sense, the resulting state would be ``maximally'' different from a pair of Bell states.  Furthermore, the amount of entanglement between $A_3B_2$ and $A_4B_1$ is given by the von Neumann entropy
\begin{equation}
S[\rho_{A_3B_2}] 
= -\cos^2(\theta_1 - \theta_2)\log_2\cos^2(\theta_1 - \theta_2) - \sin^2(\theta_1 - \theta_2)\log_2\sin^2(\theta_1 - \theta_2),
\end{equation}
where $\rho_{A_3B_2} = {\rm tr}_{A_4B_1}(|\bar{\chi}^{00}\rangle_{A_3A_4B_1B_2}\langle\bar{\chi}^{00}|)$.  Clearly, $S[\rho_{A_3B_2}]$ has maximum value $1$ when $\theta_1 - \theta_2 = \pi/4$.  Imposing these conditions, we obtain
\begin{eqnarray}
|\chi^{00}\rangle_{A_3A_4B_1B_2} & = & \frac{1}{2\sqrt{2}} 
(|0000\rangle - |0011\rangle - |0101\rangle + |0110\rangle \nonumber \\
& + & |1001\rangle + |1010\rangle + |1100\rangle + |1111\rangle)_{A_3A_4B_1B_2}.
\end{eqnarray}
From Eq.(18), we can generate a basis of 16 orthonormal states either by applying $\sigma^i$ and $\sigma^j$ to $A_3$ and $A_4$ respectively [as in Eq.(13)], or to $A_3$ and $B_1$ respectively, since $A_3B_1$ and $A_4B_2$ are maximally entangled too.  However, we cannot generate the desired basis by applying $\sigma^i$ and $\sigma^j$ to $A_3$ and $B_2$ respectively, since $A_3B_2$ and $A_4B_1$ are not maximally entangled.  Instead, we may have for instance the following orthonormal basis:
\begin{equation}
\{(\sigma^0_{A_3} \otimes \sigma^j_{B_2})|\chi^{00}\rangle_{A_3A_4B_1B_2},\ 
  (\sigma^3_{A_3} \otimes \sigma^j_{B_2})|\chi^{00}\rangle_{A_3A_4B_1B_2}\}
\end{equation}
for a $8$-dimensional subspace.  If we consider ${\cal E}_0$ for an arbitrary two-qubit state via $A_3B_2$ to $A_4B_1$, the state of particles $A_4B_1$ conditioned on Alice's measurement result $ij$:
\begin{eqnarray}
& & \frac{1}{\sqrt{p_{ij}}}
{_{A_1A_2A_3B_2}}\langle\Pi^{ij}|(|\Psi\rangle_{A_1A_2} \otimes |\chi^{00}\rangle_{A_3A_4B_1B_2}) \nonumber \\
& = & \frac{1}{\sqrt{p_{ij}}}
{_{A_1A_2A_3B_2}}\langle\Pi^{00}|
[(\sigma^i_{A_1} \otimes \sigma^j_{A_2})|\Psi\rangle_{A_1A_2} \otimes |\chi^{00}\rangle_{A_3A_4B_1B_2}],
\end{eqnarray}
where it follows from Eq.(13):
\begin{eqnarray}
|\Pi^{00}\rangle_{A_1A_2A_3B_2} & = & \frac{1}{2\sqrt{2}}(|0000\rangle + |0011\rangle - |0101\rangle + |0110\rangle \nonumber \\
& + & |1001\rangle + |1010\rangle - |1100\rangle + |1111\rangle)_{A_1A_2A_3B_2},
\end{eqnarray}
which together with Eq.(18) yield [in contrast to Eq.(15)],
\begin{eqnarray}
{_{A_1A_2A_3B_2}}\langle\Pi^{00}|\chi^{00}\rangle_{A_3A_4B_1B_2} = \frac{1}{4} & [ &
|\Psi^0_{Bell}\rangle_{A_4B_1} \times {_{A_1A_2}}\langle\Psi^0_{Bell}| + 
|\Psi^1_{Bell}\rangle_{A_4B_1} \times {_{A_1A_2}}\langle\Psi^1_{Bell}| \nonumber \\
& + & 
|\Psi^1_{Bell}\rangle_{A_4B_1} \times {_{A_1A_2}}\langle\Psi^2_{Bell}| + 
|\Psi^0_{Bell}\rangle_{A_4B_1} \times {_{A_1A_2}}\langle\Psi^3_{Bell}|]. \nonumber \\
\end{eqnarray}
This implies that faithful teleportation is possible only for partially unknown entangled states such as $\alpha_0|\Psi^0_{Bell}\rangle_{A_1A_2} + \alpha_1|\Psi^1_{Bell}\rangle_{A_1A_2}$.  From hereon, we shall focus our analysis on $|\chi^{00}\rangle_{A_3A_4B_1B_2}$.

By construction, there is absolutely zero entanglement between any one particle and any other particle.  The entanglement is purely between pairs of particles: $A_3A_4$ and $B_1B_2$, $A_3B_1$ and $A_4B_2$, and, $A_3B_2$ and $A_4B_1$.  This is in contrast to two Bell pairs where the maximal entanglement between $A_3A_4$ and $B_1B_2$ is due to those between $A_3(A_4)$ and $B_1(B_2)$.  The behavior of the entanglement associated with $|\chi^{00}\rangle_{A_3A_4B_1B_2}$ under particle loss resembles that of a GHZ state \cite{Greenberger, Dur}, in that:
\begin{equation}
S[\sigma] = 1,
\end{equation}
where $\sigma$ is the resultant density operator from partial tracing $|\chi^{00}\rangle_{A_3A_4B_1B_2}$ over any one of the four particles; i.e. the lost particle is in a completely mixed state.  Incidentally, one can teleport perfectly an arbitrary qubit from any one party to any other party if the other two parties choose to cooperate as in the teleportation protocol of Karlsson {\em et al.} \cite{Karlsson}, which employs a GHZ channel:
\begin{eqnarray}
& & |\psi\rangle_{A_1} \otimes |\chi^{00}\rangle_{A_2B_1B_2B_3} \nonumber \\
& = & \frac{1}{4}|\Psi^0_{Bell}\rangle_{A_1A_2}
[a(|000\rangle - |011\rangle - |101\rangle + |110\rangle) + b(|001\rangle + |010\rangle + |100\rangle + |111\rangle)]_{B_1B_2B_3} 
\nonumber \\
& + & \frac{1}{4}|\Psi^1_{Bell}\rangle_{A_1A_2}
[a(|001\rangle + |010\rangle + |100\rangle + |111\rangle) + b(|000\rangle - |011\rangle - |101\rangle + |110\rangle)]_{B_1B_2B_3}
 \nonumber \\
& + & \frac{1}{4}|\Psi^2_{Bell}\rangle_{A_1A_2}
[a(|001\rangle + |010\rangle + |100\rangle + |111\rangle) - b(|000\rangle - |011\rangle - |101\rangle + |110\rangle)]_{B_1B_2B_3}
 \nonumber \\
& + & \frac{1}{4}|\Psi^3_{Bell}\rangle_{A_1A_2}
[a(|000\rangle - |011\rangle - |101\rangle + |110\rangle) - b(|001\rangle + |010\rangle + |100\rangle + |111\rangle)]_{B_1B_2B_3}
. \nonumber \\
\end{eqnarray}
In particular, if $B_1$ and $B_2$ measure in the $\{|0\rangle,\ |1\rangle\}$ basis, and together with Alice communicate classically their measurement results to $B_3$, he would be able to obtain $|\psi\rangle_{B_3}$.  It is not difficult to see that the protocol works because measurements in the $\{|0\rangle,\ |1\rangle\}$ basis carried out by any two parties on $|\chi^{00}\rangle$ establish a Bell channel across the other two parties.

We should point out that in contrast to a GHZ state, $\sigma$ is entangled.  Specifically, if particle $A_3$ is lost, the nonzero negativity \cite{Vidal} between $A_4$ and $B_1B_2$ is equal to that between $B_1$ and $A_4B_2$.  This is surprising because the original entanglement was between the pairs of particles, yet it is not completely destroyed due to particle loss.  In this sense, the behavior of the entanglement associated with $|\chi^{00}\rangle_{A_3A_4B_1B_2}$ under particle loss also resembles that of a W state \cite{Zeilinger, Dur}.  However, a further particle loss will destroy all entanglement.

Lastly, $|\chi^{00}\rangle$ truly differs from the GHZ and W states in that both these states do not enable the teleportation of an arbitrary two-qubit state.  Indeed, they are SLOCC inequivalent (see Ref.\cite{Verstraete}).  $|\chi^{00}\rangle$ is a ``new'' genunie multipartite entangled state.  Note that we are not claiming that $|\chi^{00}\rangle$ is LOCC inequivalent to either the GHZ or W state.  This would require further work.  For now, we shall turn our attention to dense coding \cite{Bennett2}.

A dense coding scheme ${\cal D}_0$ using $|\chi^{00}\rangle_{A_3A_4B_1B_2}$, which ``mirrors'' ${\cal E}_0$ is the following.  $A_3$ and $A_4$ encode their message using $\sigma^i_{A_3}$ and $\sigma^j_{A_4}$, and send their particles to $B_1$ and $B_2$ respectively.  $B_1$ and $B_2$ will then decode the message by performing a joint measurement on all four particles in the $\{|\chi^{ij}\rangle_{A_3A_4B_1B_2}\}$ basis.  It is easy to see that ${\cal D}_0$ works perfectly, enabling $A_3$ and $A_4$ to communicate 4 bits of classical information with $B_1B_2$ by sending in total 2 particles.  This is impossible with a four-party GHZ or W state.  However, we note that whereas $A_3$ and $A_4$ may encode their message locally and hence independently, $B_1$ and $B_2$ are compelled to read the message together.  One is not able to do it without the other's presence and cooperation.  This is in contrast to a straightforward extension of the original dense coding scheme of Bennett {\em et al.} \cite{Bennett2} to one involving two Bell states shared between $A_3(A_4)$ and $B_1(B_2)$, where $B_1$ and $B_2$ can individually read the respective message from $A_3$ and $A_4$.  We denote this scheme by ${\cal S}_0$.  This difference between ${\cal D}_0$ and ${\cal S}_0$ lies in the maximal entanglement between $A_3B_1$ and $A_4B_2$.  In terms of the numbers of particles sent and the amount of classical information communicated, both ${\cal D}_0$ and ${\cal S}_0$ are exactly the same:
\begin{equation}
4 = \log_22^4 = 4\log_22 = 2\log_22 + 2\log_22 = 2 + 2.
\end{equation}
An immediate example of a situation where ${\cal D}_0$ could have an advantage over ${\cal S}_0$ is the following: $A_3$ and $A_4$ wish to send some message to both $B_1$ and $B_2$, which they must both read at the same time together regardless of whether $A_3$ or $A_4$'s particle reaches $B_1$ or $B_2$ first.  We note that ${\cal D}_0$ works equally well between $A_3B_1$ and $A_4B_2$, but not between $A_3B_2$ and $A_4B_1$ because the entanlgement between them is not maximal.  In fact, from Eq.(19), we see that only
\begin{equation}
3 = \log_22^3 = 3\log_22 = \log_22 + 2\log_22 = 1 + 2
\end{equation}
bits of information can be transferred, if $A_3$ cooperate with $B_2$ by only encoding her qubit with either $\sigma^0$ or $\sigma^3$.  In this case, $A_4B_1$ decode by measuring in the basis, Eq.(19), for a 8-dimensional subspace.  This is consistent with Eq.(22).

In conclusion, we have shown that faithful teleportation of an arbitrary two-qubit state and dense coding are possible with $|\chi^{00}\rangle$.  These can similarly be achieved using two Bell pairs.  However, by construction, this state is different from a pair of Bell states.  It is a genuine four-partite entangled state, which has properties that differ from those of four-party GHZ and W states.  It could play an analogous role to $|\Psi^0_{Bell}\rangle$ in the theory of multipartite entanglement.


\begin{thebibliography}{99}
\bibitem{Nielsen} M. A. Nielsen and I. L. Chuang, {\em Quantum Computation and Quantum Information} (Cambridge University Press, Cambridge, 2000).
\bibitem{Bennett1} C. H. Bennett, G. Brassard, C. Crepeau, R. Josza, A. Peres, and W. K. Wootters, Phys. Rev. Lett. {\bf 70}, 1895 (1993).
\bibitem{Hill} S. Hill and W. K. Wootters, Phys. Rev. Lett. {\bf 78}, 5022 (1997).
\bibitem{Wootters} W. K. Wootters, Phys. Rev. Lett. {\bf 80}, 2245 (1998).
\bibitem{Horodecki} M. Horodecki, P. Horodecki and R. Horodecki, Phys. Rev. A {\bf 60}, 1888 (1999).
\bibitem{Lee} J. Lee, H. Min and S. D. Oh, Phys. Rev. A {\bf 66}, 052318 (2002). 
\bibitem{Rigolin} G. Rigolin, Phys. Rev. A {\bf 71}, 032303 (2005).
\bibitem{Braunstein} S. L. Braunstein, G. M. D'Ariano, G. J. Milburn, and M. F. Sacchi, Phys. Rev. Lett. {\bf 84}, 3486 (2000).
\bibitem{Greenberger} D. M. Greenberger, M. A. Horne and A. Zeilinger, in {\em Bell's Theorem, Quantum Theory, and Conceptions of the Universe}, edited by M. Kafatos (Kluwer Academic, Dordrecht, 1989), pp. 73 - 76.
\bibitem{Dur} W. D\"ur, Phys. Rev. {\bf A 63}, 020303 (2001).
\bibitem{Karlsson} A. Karlsson and M. Bourennane, Phys. Rev. A {\bf 58}, 4394 (1998).
\bibitem{Vidal} G. Vidal and R. F. Werner, Phys. Rev. A {\bf 65}, 032314 (2002).
\bibitem{Zeilinger} A. Zeilinger, M. A. Horne and D. M. Greenberger, NASA Conf. Publ. No. 3135 (National Aeronautics and Space Administration, Code NTT, Washington, DC, 1997).
\bibitem{Verstraete} F. Verstraete, J. Dehaene, B. De Moor and H. Verschelde, Phys. Rev. {\bf A 65}, 052112 (2002).
\bibitem{Bennett2} C. H. Bennett and S. J. Wiesner, Phys. Rev. Lett. {\bf 69}, 2881 (1992).
\end{thebibliography}
\end{document}